\documentstyle[epsfig]{mn}

\newif\ifAMStwofonts
\AMStwofontstrue

\newcommand{\be}{\begin{equation}}    
\newcommand{\ee}{\end{equation}}
\newcommand{\beq}{\begin{eqnarray}}
\newcommand{\eeq}{\end{eqnarray}}
\newcommand{\f}[2]{\frac{#1}{#2}}
\newcommand{\vol}[1]{{\bf #1}}

\def\op{ \ $ }
\def\cl{$ \ }
\def\nn{\nonumber}

\ifoldfss
  \ifCUPmtlplainloaded \else
    \NewTextAlphabet{textbfit} {cmbxti10} {}
    \NewTextAlphabet{textbfss} {cmssbx10} {}
    \NewMathAlphabet{mathbfit} {cmbxti10} {} 
    \NewMathAlphabet{mathbfss} {cmssbx10} {} 
  \fi
  \ifAMStwofonts
    \ifCUPmtlplainloaded \else
      \NewSymbolFont{upmath} {eurm10}
      \NewSymbolFont{AMSa} {msam10}
      \NewMathSymbol{\upi}     {0}{upmath}{19}
      \NewMathSymbol{\umu}     {0}{upmath}{16}
      \NewMathSymbol{\upartial}{0}{upmath}{40}
      \NewMathSymbol{\leqslant}{3}{AMSa}{36}
      \NewMathSymbol{\geqslant}{3}{AMSa}{3E}

      \let\geq=\geqslant 
    \fi
  \fi
\fi 

\ifnfssone
  \newmathalphabet{\mathit}
  \addtoversion{normal}{\mathit}{cmr}{m}{it}
  \addtoversion{bold}{\mathit}{cmr}{bx}{it}
  \newmathalphabet{\mathbfit} 
  \addtoversion{normal}{\mathbfit}{cmr}{bx}{it}
  \addtoversion{bold}{\mathbfit}{cmr}{bx}{it}
  \newmathalphabet{\mathbfss} 
  \addtoversion{normal}{\mathbfss}{cmss}{bx}{n}
  \addtoversion{bold}{\mathbfss}{cmss}{bx}{n}
  \ifAMStwofonts
    \ifCUPmtlplainloaded \else
      \UseAMStwoboldmath
      \makeatletter
      \new@mathgroup\upmath@group
      \define@mathgroup\mv@normal\upmath@group{eur}{m}{n}
      \define@mathgroup\mv@bold\upmath@group{eur}{b}{n}
      \edef\UPM{\hexnumber\upmath@group}
      \new@mathgroup\amsa@group
      \define@mathgroup\mv@normal\amsa@group{msa}{m}{n}
      \define@mathgroup\mv@bold\amsa@group{msa}{m}{n}
      \edef\AMSa{\hexnumber\amsa@group}
      \makeatother
      \mathchardef\upi="0\UPM19
      \mathchardef\umu="0\UPM16
      \mathchardef\upartial="0\UPM40
      \mathchardef\leqslant="3\AMSa36
      \mathchardef\geqslant="3\AMSa3E

      \let\geq=\geqslant 
    \fi
  \fi
\fi 

\ifnfsstwo
  \DeclareMathAlphabet{\mathbfit}{OT1}{cmr}{bx}{it}
  \SetMathAlphabet\mathbfit{bold}{OT1}{cmr}{bx}{it}
  \DeclareMathAlphabet{\mathbfss}{OT1}{cmss}{bx}{n}
  \SetMathAlphabet\mathbfss{bold}{OT1}{cmss}{bx}{n}
  \ifAMStwofonts
    \ifCUPmtlplainloaded \else
      \DeclareSymbolFont{UPM}{U}{eur}{m}{n}
      \SetSymbolFont{UPM}{bold}{U}{eur}{b}{n}
      \DeclareSymbolFont{AMSa}{U}{msa}{m}{n}
      \DeclareMathSymbol{\upi}{0}{UPM}{"19}
      \DeclareMathSymbol{\umu}{0}{UPM}{"16}
      \DeclareMathSymbol{\upartial}{0}{UPM}{"40}
      \DeclareMathSymbol{\leqslant}{3}{AMSa}{"36}
      \DeclareMathSymbol{\geqslant}{3}{AMSa}{"3E}

      \let\geq=\geqslant 
    \fi
  \fi
\fi 

\ifCUPmtlplainloaded \else
  \ifAMStwofonts \else 
    \def\upi{\pi}
    \def\umu{\mu}
    \def\upartial{\partial}
  \fi
\fi

\title[The imprint of the equation of state on the axial
w-modes of oscillating  neutron stars]
{The imprint of the equation of state on the axial
w-modes of oscillating  neutron stars}
\author[O. Benhar, E. Berti, V. Ferrari]
       {Omar Benhar$^{1,2}$, Emanuele Berti$^2$,Valeria Ferrari$^{1,2}$\\
        $^{1}$INFN, Sezione Roma 1\\
        $^{2}$Dipartimento di Fisica ``G. Marconi",
Universit\'a degli Studi di Roma, ``La Sapienza",\\ P.le A. Moro
2, 00185 Roma, Italy}

\date{December  1998}

\pagerange{\pageref{firstpage}--\pageref{lastpage}}
\pubyear{1998}
\begin{document}

\maketitle
\label{firstpage}

\begin{abstract}

We discuss the dependence of the pulsation frequencies of 
the axial quasi-normal modes of a nonrotating neutron star upon the 
equation of state describing the star interior.
The continued fraction method has been used to compute the complex
frequencies for a set of equations of state based on different physical 
assumptions and spanning a wide range of stiffness.
The numerical results show that the detection of axial gravitational
waves would allow to discriminate between the models underlying the 
different equation of states, thus providing relevant information on 
both the structure of neutron star matter and the nature of the 
hadronic interactions.

\end{abstract}

\begin{keywords}
gravitational waves  -- equation of state -- 
quasi-normal modes of compact stars.
\end{keywords}

\section{Introduction}
The search for gravitational waves has received a big burst during the
last decade.  In addition to the already operating resonant antennas
(EXPLORER, AURIGA, NAUTILUS, ALLEGRO, NIOBE), 
ground-based interferometers (VIRGO, LIGO, GEO600, TAMA) will be soon 
operating, and a future space-based interferometer,  LISA, will extend
the  exploration to low frequencies, unaccessible to  other detectors.
Thus, the detection of gravitational waves  is close to become reality.
Among the astrophysical processes that are associated with emission 
of gravitational waves,
the gravitational collapse and the coalescence of compact bodies
are likely to be the most efficient  sources.
Both processes leave behind a compact body, either a neutron star 
or a black hole, which is expected to radiate the mechanical energy 
residual of its violent birth through gravitational waves. 
These waves will be emitted at frequencies and with 
damping times characteristic of the quasi-normal modes of the source.
It is interesting to note that these modes may be excited also in 
other astrophysical processes that can be observed.
For instance, stellar oscillations are thought to be at the origin of
the drifting subpulses and micropulses detected in some
radio sources, and of the quasi-periodic variability seen in 
some X-ray burst sources and in a number of bright X-ray sources
(McDermott, Van Horn \& Hansen 1988).

The complex frequencies of the quasi-normal modes
carry information on the internal structure
of the emitting source. For black holes, it has been shown
that they depend
exclusively on the parameters that identify the spacetime geometry: 
the mass, charge and angular momentum.
For stars, the situation is far less simple, since the
quasi-normal mode eigenfrequencies depend on the equation of
state (EOS) prevailing in the interior, on which not much is known.  
It is therefore interesting to compute these frequencies 
for different EOS's proposed to describe matter at supernuclear
densities, and explore the possibility of extracting from them
information on the internal structure of the star.

The quasi-normal mode frequencies can be computed by studying the 
source-free, adiabatic 
perturbations of an equilibrium configuration with an assigned 
EOS, and by solving the linearized Einstein
equations, coupled with the equations of hydrodynamics,
with suitably posed boundary conditions.

If the unperturbed star is assumed to be
static and non rotating, it is convenient
to expand all perturbed tensors in tensorial spherical harmonics,
and since these harmonics have a different
behaviour  under the angular transformation
\op \theta\rightarrow\pi-\theta,\cl
\op \varphi\rightarrow\pi+\varphi,\cl
the separated equations split in two decoupled sets:
the {\it polar} or {\it even},
belonging to the parity \op (-1)^{\ell},\cl
and the {\it axial} or {\it odd}, belonging to the parity
\op (-1)^{(\ell+1)}.\cl
The polar equations  are the relativistic generalization of the
tidal perturbations of newtonian theory, and couple the  
perturbations of the gravitational field
with the perturbations of the fluid composing the star. 
As in newtonian theory, the classification of the polar modes
is based on the behaviour of the perturbed fluid according to the
restoring force that is prevailing: the g-modes, or gravity
modes, when the force is due to the eulerian change in the density, 
the p-modes, when it is due to a change in pressure, and 
the f-mode, that  is the
generalization of the only possible mode of oscillation of an
incompressible homogeneous sphere (Chandrasekhar 1964).
This classification scheme was introduced by Cowling
(Cowling 1941).
In addition to the  g, f, p modes, that exist also in
newtonian theory,  in general relativity there exists a
new family of modes that are essentially spacetime modes,
since the corresponding motion of the fluid is negligible 
(Kokkotas \& Schutz 1992).
They are named w-modes, and
are characterized by frequencies typically higher than those of 
the g, f, p modes, and much smaller damping times, i.e. these modes are
highly damped. 

The  frequency of the f-mode was determined by
Lindblom and Detweiler  for several EOS's (Lindblom \& Detweiler 1983),
and more recently this study has been extended to include a few modern
equations of state suggested for neutron stars, for which also
the frequencies of the first p-mode and of the fundamental polar w-mode
have been computed. In addition, it has been 
shown that it is possible to infer empirical relations between the mode 
frequency and the macroscopic parameters of the star: the mass and the radius
(Andersson \& Kokkotas 1998).

Unlike the polar perturbations, the axial perturbations are 
not coupled with fluid motion.
They do not have a newtonian counterpart, and their radial 
evolution is described by
a Schr\"odinger-like equation with a potential barrier that
depends on the distribution of energy and pressure in the
interior of the star in the equilibrium configuration
(Chandrasekhar \& Ferrari 1991a). Thus,  the EOS of the fluid
has the sole  role of determining the shape of the potential 
inside the star.

The axial quasi-normal modes divide in two classes:
the w-modes, highly damped and with properties
similar to the polar w-modes, and the
slowly damped s-modes (Chandrasekhar \& Ferrari 1991b).
These modes appear if the star is extremely compact, so that the potential
well in the interior becomes deep enough to allow for
the existence of one or more quasi-stationary states,
i.e. of  quasi normal modes.

In this paper  we shall calculate the frequencies of the axial
quasi-normal modes for the following  EOS's;  
whenever possible, we follow the classification scheme introduced 
by Arnett \& Bowers (1977):
\begin{itemize}
\item{} Pandharipande, model A. The star is described
in terms of pure neutron matter, the dynamics of which is dictated by a
nonrelativistic hamiltonian containing a 
semi-phenomenological interaction potential (Pandharipande 1971a). 
\item{} Pandharipande, model B. A generalization of model A including 
the appearance of protons, electrons and muons in \op \beta$-equilibrium, 
as well as of heavier baryons (hyperons and nucleon resonances), at 
sufficiently high densities (Pandharipande 1971b).
\item{} Wiringa, Fiks and Fabrocini, model WFF. Neutron star matter is
treated as a mixture of neutrons, protons, electrons and muons 
in \op \beta$-equilibrium. The nuclear hamiltonian includes 
two- and three-body potentials. With respect to models A and B, the ground 
state energy of neutron star matter is computed using a more sophisticated 
and accurate many-body technique (Wiringa, Fiks \& Fabrocini 1988). 
\item{} Akmal, Pandharipande and Ravenhall, model APR1. Similar to 
model WFF, but uses state of the art parametrizations of both the two- and 
three-body potentials (Akmal, Pandhripande \& Ravenhall 1998).
\item{}  Akmal, Pandharipande and Ravenhall, model APR2. Includes selected 
relativistic corrections to model APR1 (Akmal {\it et al.} 1998).
\item{} Pandaripande \& Smith 1975, model L. Neutrons are assumed to
interact through exchange of \op \omega-$, \op \rho-$ and \op
\sigma-$mesons. While the exchange of heavy particles ($\omega \cl and
\op \rho$) is described in terms of nonrelativistic potentials, the
effect of the \op \sigma-$meson is taken into account using relativistic
field theory and the mean-field approximation (Pandharipande \& Smith 1975). 
The physical assumptions underlying this {\em hybrid} approach are somewhat 
different from those of the previous models.
\end{itemize}
A useful way of classifying EOS's is through
their stiffness, which can be quantified in terms of the speed of sound
\op v_s$: stiffer EOS's correspond to higher sound speeds. The stiffness
is also a measure of compressibility: stiffer EOS's
correspond to more incompressible matter.

The EOS's considered in this paper can then be ordered according to 
increasing stiffness as follows:   B, A, 
APR2, APR1, WFF, L.

Though the s-modes have been shown to exist for homogeneous stars
\footnote{
It should be mentioned that \op\epsilon=const$, $\epsilon$ being 
the energy density, is the stiffest EOS, since \op v_s=\infty$. 
}
with high enough compactness, the EOS's we have considered do not
lead to  potential wells deep enough to allow for the appearance of
any s-mode.  
Thus we have focussed on the imprint that different EOS's leave
on the axial w-modes. In particular, we have computed the  pulsation
frequency and the damping time of the fundamental w-mode for the
aforementioned EOS's, by using the continued fraction
method, first developed by Leaver to find the 
quasi-normal mode frequencies of black holes (Leaver 1985, 1986), 
and which we have suitably modified to make it applicable to the present 
context.

We shall show that, as for the polar w-modes, it is possible to 
derive empirical relations that allow to
determine the mass and radius of the star.  
However, the most remarkable result is 
that the range within which the pulsation frequencies vary
is peculiar of each EOS, much more than in the polar case.

The plan of the paper is the following. In section 2 we shall 
write the equations governing the axial perturbations of a non rotating star.
In section 3 the EOS's will be shortly described and 
compared.
In section 4 the continued fraction method will be outlined
and section 5 will be devoted to a discussion of the results.

\section{A Schr\"odinger equation for the axial perturbations}\label{schrod}

The equations for the axial perturbations of a non rotating star
can be considerably simplified by introducing, after separating
the variables, a  function \op Z_\ell(r),\cl
constructed from the radial part of the axial metric
components, where \op\ell\cl is the harmonic index
(Chandrasekhar \& Ferrari 1991a). 
\op Z_\ell(r)\cl satisfies the following equation
\beq
\label{zz1}
&&\frac{d^{2}Z_\ell}{ dr_{*}^{2}}+
[\omega^{2}-V_\ell(r)]Z_\ell=0,\\\nn
&&V_{\ell}(r)=\frac{e^{2\nu}}{ r^{3}}[l(l+1)r+r^{3}(\epsilon-p)-6m(r)],
\eeq
where
\op
r_*=\int_0^r e^{-\nu+\mu_2}dr,
\cl
and \op\epsilon\cl and \op p\cl are the energy density and the pressure of
the perfect fluid composing the star.
The function \op\nu(r)\cl appearing in the potential barrier,
can be found by integrating the equation
\be
\nu_{,r}=-\frac{p_{,r}}{\epsilon +p},
\ee
and imposing the boundary condition that, at the surface of the star,
the metric reduces to the Schwarzschild metric, i.e. 
\op e^{2\nu(R)}=1-\frac{2M}{R},\cl where \op M = m(R) \cl is the  mass of 
the star.
Outside the star, where \op\epsilon\cl and \op p \cl vanish,
eq. (\ref{zz1}) reduces to well known Regge-Wheeler  equation 
\beq
\label{reggew}
&&\frac{d^{2}Z_\ell}{ dr_{*}^{2}}+
[\omega^{2}-V_\ell(r)]Z_\ell=0,\\\nn
&&V_{\ell }(r)=\frac{e^{2\nu}}{ r^{3}}[l(l+1)r-6M],\qquad\hbox{and}\qquad
e^{2\nu}=1-\frac{2M}{ r}.
\eeq

Thus the axial perturbations of  a star
are  fully described by a Schr\"odinger-like equation
with a potential barrier that depends on the distribution of energy-density
and pressure inside the star in its equilibrium configuration. 

The quasi-normal modes are defined to be solutions of eq. (\ref{reggew})
which satisfy the condition of being regular at \op r=0,\cl and that
behave as a purely outgoing wave at radial infinity, i.e.
\be
Z_\ell \sim e^{-i\omega r_*}\qquad   r_* \rightarrow \infty.
\ee

\section{Models of neutron star matter equation of state}

In this work, we have used six different models of neutron star matter 
EOS at supernuclear density ($\epsilon > 2.8\cdot 10^{14}$ g/cm$^3$). 
To obtain the EOS over the whole density range relevant to neutron stars, 
models A, B , WFF and L have been matched to the EOS at subnuclear 
densities obtained by Feynman, Metropolis \& Teller (1949),
Baym, Pethick \& Sutherland (1971)
and Baym, Bethe \& Pethick (1971), whereas models APR1 and APR2 
have been supplemented with the recent results by Pethick, 
Ravenhall \& Lorenz (1995). However, the details
of the EOS at subnuclear density have negligible influence on both the 
neutron star properties and the calculated pulsation frequencies.
The equations of state we consider are plotted in fig. 1 over the entire
range of density, whereas in fig. 2 the  different high density behaviours
are compared.

\begin{figure}
\begin{center}
\leavevmode
\centerline{\epsfig{figure=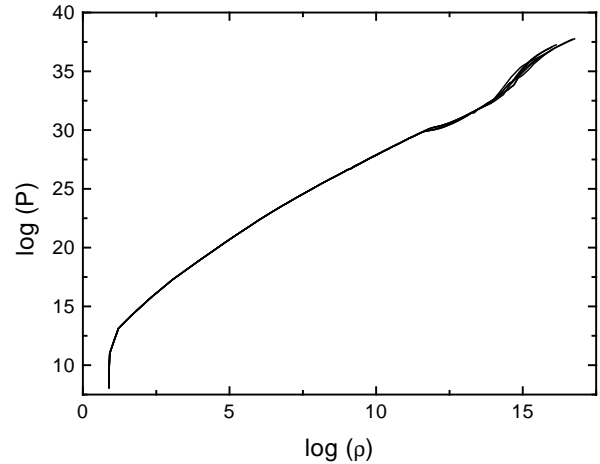,angle=000,width=10cm}}
\caption{
The logarithm of pressure (dynes$\cdot$cm$^2$) versus the logarithm of the 
energy density (g$\cdot$cm$^{-3}$) for the EOS's considered
in this paper over the whole density range. 
In the region of subnuclear densities the curves corresponding to different 
EOS's sit on top of each other.
}\label{EOSTOT}
\end{center}
\end{figure}
\begin{figure}
\begin{center}
\leavevmode
\centerline{\epsfig{figure=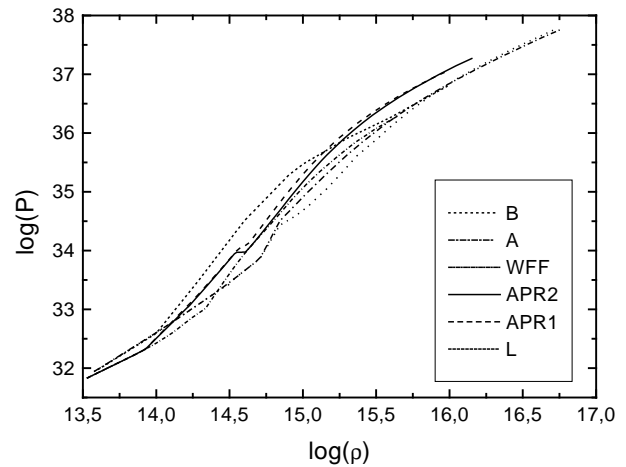,angle=000,width=10cm}}
\caption{
A blow-up of the region of supernuclear densities of figure \ref{EOSTOT}.
The units are the same as in the previous figure.  
It can be seen that, in the region relevant to neutron star properties,
the EOS's can be ordered according to increasing stiffness 
as explained in the text:
B, A, WFF, APR2, APR1, L.
}
\label{EOSZOOM}
\end{center}
\end{figure}

\begin{figure}
\begin{center}
\leavevmode
\centerline{\epsfig{figure=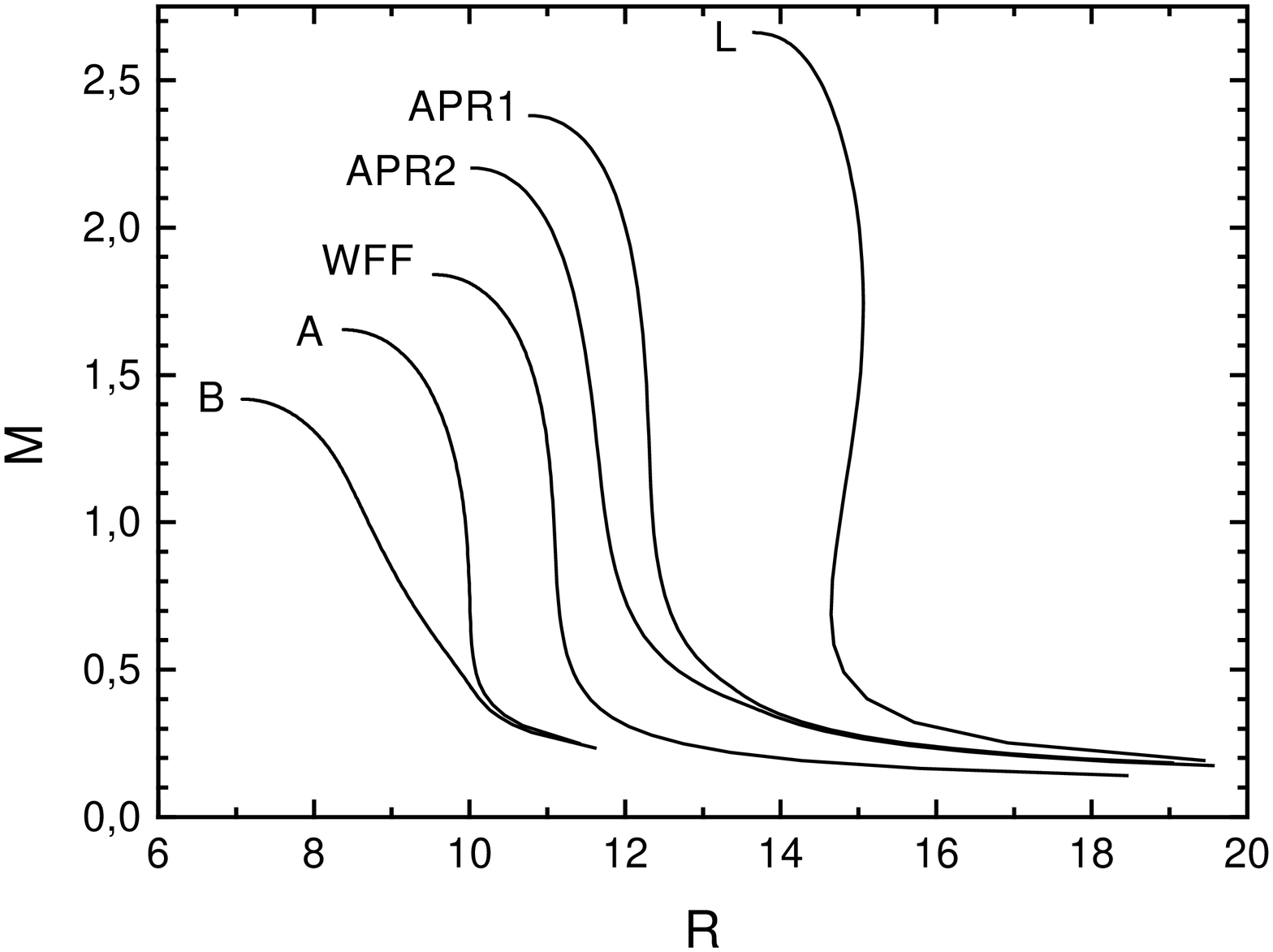,angle=000,width=10cm}}
\caption{The mass-radius relation is plotted for the EOS's considered
in this paper, in the  range of stability. $M$ is expressed in solar mass
units, and $R$ in km.}
\label{MR}
\end{center}
\end{figure}

In fig. \ref{MR} we plot the mass radius relation
for the above mentioned EOS's,
in the range where we have checked that the stellar models are stable
against radial perturbations.

The EOS's denoted by A (Pandharipande 1971a), B (Pandharipande 1971b),
L (Pandhripande {\it et al.} 1975) and WFF (Wiringa {\it et al.} 1988) 
have been widely used in the past to study equilibrium configurations 
of neutron stars 
(Arnett {\it et al.} 1977; Pandharipande, Pines \& Smith 1976; 
Wiringa {\it et al.} 1988) 
and, more recently, to compute their pulsation frequencies associated with 
emission of gravitational waves (Andersson \& Kokkotas 1998). 
On the other hand, the models
denoted APR1 and APR2 (Akmal {\it et al.} 1998) have been developed 
within the last two years and have never before been employed to study 
nonstatic neutron star properties.

With the only exception of model L, the EOS's considered have been obtained
using nonrelativistic nuclear many--body theory. Within this approach, 
nuclear matter is treated as a system of pointlike protons and 
neutrons, whose dynamics is described by the nonrelativistic hamiltonian:
\be
H = \sum_i \frac{p_i^2}{2m} + \sum_{j>i} v_{ij} + \sum_{k>j>i} V_{ijk}\ ,
\label{hamiltonian}
\ee
where \op m \cl and \op p_i \cl denote the nucleon mass 
and momentum, respectively, whereas \op v_{ij} \cl and \op V_{ijk} \cl 
describe two- and three-nucleon interactions. The two-nucleon potential is 
obtained from 
a fit to the properties of both the bound and scattering states of the 
two-nucleon system (Wiringa, Stoks \& Schiavilla 1995), while 
the three-body term \op V_{ijk} \cl has to be included in order 
to account for the binding energies of the three-nucleon bound 
states (Pudliner {\it et al.} 1995) . 

Obtaining the EOS of nuclear matter at zero temperature amounts to 
evaluate its ground state energy per particle \op E_0 \cl as a function 
of baryon density, which in turn 
requires the solution of the many--body Schr\"odinger equation with the
hamiltonian of eq.(\ref{hamiltonian}). The EOS's
considered in this work have been obtained using the Rayleigh--Ritz 
variational principle and the formalism of cluster expansions, which 
allows to write \op E_0 \cl as the sum of contributions arising from 
subsystems containing an increasing number of particles. 
In the pioneering works of Pandharipande (1971a, 1971b) the calculations 
were carried out at lowest order in the cluster expansion, 
i.e. including only two--body cluster contributions. 
The more recent WFF, APR1 and APR2 
EOS's have been obtained using the Fermi-Hyper-Netted-Chain (FHNC) 
integral equations (Fantoni \& Rosati 1975, Pandharipande \& Wiringa 1979) 
to take into account the contributions of 
the relevant terms in the cluster expansion to all orders. It has to be
pointed out that the FHNC equations allow for a very precise calculation
of the ground state expectation value of the nuclear hamiltonian. 
Moreover, within this computational scheme it is possible to use 
highly realistic trial wave functions, whose structure reflects 
the complexity of the nuclear interactions. 
In this respect, the results obtained by Wiringa {\it et al.} (1988) and by Akmal {\it et al.} (1998) are more accurate than those obtained 
by Pandharipande (1971a, 1971b).

As far as the nuclear hamiltonian is concerned, it is important to realize 
that the two--body potential models developed over the past few 
years (Wiringa {\it et al.} 1995), as the Argonne \op v_{18}\cl 
potential used in model APR1 and APR2, represent a significant improvement 
over the previously available ones. 
They are fitted to a larger and more complete database of nucleon--nucleon
scattering data and explicitly include the effect of 
charge--symmetry--breaking terms in the scattering amplitude. 
Results by Akmal {\it et al.} (1998) show that using a two--body potential 
of the last generation leads to qualitative changes in the density 
dependence of nuclear and neutron matter energy, suggesting the possibility
of a transition from the standard uniform phase to a spin--isospin 
ordered phase associated with condensation of neutral pions
(Tamiya \& Tamagaki 1988, Benhar 1981). 

We will now briefly summarize the main features of the model EOS's 
used in this work. 
In the case of EOS A (Pandharipande 1971a) neutrons are allowed to 
interact through a two--body potential, the three--body potential is 
neglected and the cluster expansion is truncated at the two--body level.
The hamiltonian has been constructed using an early  model of 
\op v_{ij}$, developed by Reid in the late sixties (Reid 1968). 

A more realistic description of neutron star matter requires inclusion 
of the effect of inverse $\beta$-decay, which eventually leads to the 
appearance of a mixture of neutrons, protons, electrons and possibly 
muons in $\beta$--equilibrium. Models B, WFF, APR1 and APR2 
are all based on this picture. As already mentioned, the calculations by 
Wiringa {\it et al.} (1988) and Akmal {\it et al.} (1998) have been 
carried out using the FHNC equations. 
Among the different EOS's discussed in the first reference, we have used 
the one corresponding to the so called Urbana \op v_{14} \cl  two--body 
potential, supplemented by a purely phenomenological three--nucleon 
interaction.

From the computational point of view, the EOS of model 
APR1 (Akmal {\it et al.} 1998) has been obtained using basically the 
same approach as 
Wiringa {\it et al.} (1988). 
The main difference between models APR1 and WFF comes from the nuclear 
hamiltonians. 
Model APR1 uses the new Argonne $v_{18}$ potential and a three--body 
interaction, referred to as Urbana IX, featuring an attractive 
component, due to pion--exchange processes with excitation of a 
\op \Delta \cl resonance in the intermediate state, and a phenomenological
repulsive part at shorter range. 

The relevance of relativistic corrections to the standard nuclear 
many--body picture has been studied in model APR2 
(Akmal {\it et al.} 1998) by including boost corrections to the 
nucleon--nucleon interaction of order $(v/c)$, $v$ being the 
velocity of the center of mass of the interacting pair.
In order to  fit the observed  properties of the three--nucleon systems
after the inclusion of relativistic corrections, 
the change in the two--body part of the interaction 
requires the three--body potential to be modified accordingly.
The resulting three-body interaction is referred to as UIX$^*$.

The main problem associated with models A, B, WFF, APR1 and APR2 is 
that they lead to a violation of causality, since the predicted 
speed of sound exceeds the speed of light at very high density. 
However, this pathology only occurs at 
densities much greater than those relevant to the study of 
neutron star structure.

The above problem can be avoided treating neutron star matter within
relativistic field theory, with a lagrangian describing the nuclear 
interaction in terms of one--boson exchange processes. 
The coupling constants are chosen in such a way as to reproduce selected 
nuclear properties, and the corresponding field equations are solved 
in the mean--field approximation.
Following the pioneering work of Walecka (1974) this approach has been 
used by many authors. 
Model L (Pandharipande {\it et al.} 1975) includes the exchange of 
two vector mesons ($\omega$ and $\rho$) and one scalar meson ($\sigma$). 
It can be regarded as a somewhat {\it hybrid} model, since only the 
$\sigma$-exchange interaction is treated within the mean--field 
approximation, whereas the exchange of the heavier vector mesons is 
described by static potentials. 

The EOS's employed in this work can be classified according to their
stiffness, providing a measure of the incompressibility of neutron star
matter. Comparison between models A and B shows that the appearance of heavy
baryons leads to a significant softening of the EOS at high density. This
behaviour can be easily understood noting that, due to their low
concentrations, the heavy baryons carry a kinetic energy much smaller
than that carried by the neutrons. Hence, replacing neutrons with 
heavy baryons produces a sizeable decrease of the pressure.

In the case of neutron matter, either pure or in $\beta$-equilibrium, the
stiffness of the EOS is mainly determined by the nuclear hamiltonian. The 
EOS's of models A, WFF, APR1 and APR2, in which the nuclear hamiltonian 
is contrained by nucleon--nucleon scattering data, are rather close to
each other. On the other hand, using a hamiltonian that does not reproduce 
nucleon--nucleon data, as in model L, may result in a significantly 
difference in stiffness.

It is interesting to note that the comparison between models APR1 and APR2 
suggests that the inclusion of relativistic corrections leads to a 
softening of the EOS. Hence, the fact that model L provides the stiffest 
EOS should be ascribed to the different dynamical assumptions, rather than
to its relativistic nature.

\section{The continued fraction method}

As stated in section \ref{schrod}, the axial quasi--normal modes
of a pulsating star are solutions of eq.(\ref{zz1}) satisfying 
the boundary conditions imposed by physical requirements:
\op Z_\ell \cl should be regular at the origin, and have the behaviour 
of a purely outgoing wave at infinity.
To compute the mode frequencies one has to find the (complex) values,
\op \omega=\omega_0+i\omega_i \cl, for which these boundary conditions 
are satisfied.

The numerical determination of the
quasi--normal mode frequencies is non--trivial, especially 
for modes with large imaginary parts (strongly damped modes). 
The reason is simple to understand. Solutions of eq.(\ref{zz1}) 
representing outgoing and ingoing waves at infinity
have the asymptotic behaviour
\be
Z_\ell^{out} \sim e^{r_*/\tau} \qquad \hbox{and} \qquad Z_\ell^{in} \sim e^{-r_*/\tau}
\ee
as \op r_*\to \infty$, where  \op \tau=1/\omega_i \cl is the damping time.
Therefore, identifying by numerical integration the purely outgoing solutions (that is, those solutions for which \op Z_\ell^{in} \cl is zero) becomes increasingly difficult as the damping 
of the mode increases.

The same problem occurs also in the case of quasi--normal modes of black holes,
and was solved by Leaver (Leaver 1985, 1986). 
He found a continued fraction relation that can be
regarded as an implicit equation which identifies the quasi--normal frequencies, thus circumventing the need to perform an integration out to big values of \op r_*$.
This method was subsequently reformulated and applied to 
the  polar oscillations 
of a star (Leins, Nollert \& Soffel, 1993).
Here we show that it can be used also in the case of {\it axial} perturbations.

In this section we will use dimensionless geometrical units, \op c=G=2M=1$.
In these units the Regge--Wheeler equation, which describes the perturbed
spacetime outside the star, becomes:
\be\label{mastereq}
\f{d^2 Z_\ell}{dr_*^2}+\left[\omega^2-\left(1-\f{1}{r}\right)
\left(\f{\ell(\ell+1)}{r^2}-\f{3}{r^3}\right)\right]Z_\ell=0,
\ee
and
\be
r_*=r+\ln(r-1).
\ee
We shall now write the solution of the Regge-Wheeler equation in a
power-series form as follows.
Defining \op v\equiv 1-\f{a}{r}, \cl where \op r=a\cl is some point
outside the star,
and introducing a function \op \phi_\ell(v)$, related to \op Z_\ell(r)\cl by:
\be\label{subst}
Z_\ell(r)=(r-1)^{-i\omega}e^{-i\omega r}\phi_\ell(v)\equiv \chi(r)\phi_\ell(v),
\ee
one finds that \op \phi_\ell\cl satisfies the differential equation:
\beq\label{stellLeav}
&&(c_0+c_1v+c_2 v^2+c_3 v^3){\phi_\ell}_{,vv}+ \\
\nn
&&+(d_0+d_1 v+d_2 v^2){\phi_\ell}_{,v}+(e_0+e_1 v)\phi_\ell=0.
\eeq
The constants depend only on \op \omega$,\op \ell \cl and \op a \cl 
through the relations:
\beq\label{coefficienti}
\nn
c_0=1-\f{1}{a} \qquad c_1=\f{3}{a}-2 \qquad c_2=1-\f{3}{a} 
\qquad c_3=\f{1}{a}\\
\nn
d_0=-2i\omega a+\f{3}{a}-2 \qquad d_1=2\left(1-\f{3}{a}\right) 
\qquad d_2=\f{3}{a}\\
\nn
e_0=\f{3}{a}-\ell(\ell+1) \qquad e_1=-\f{3}{a}.
\eeq
Let us now perform a power expansion of $\phi_\ell(v)$:
\be\label{expans}
\phi_\ell(v)=\sum_{n=0}^\infty a_n v^n.
\ee
By substituting this expression in eq.(\ref{stellLeav}),  the 
expansion coefficients \op a_n\cl are found to
satisfy a four--term recurrence relation 
of the form:
\be\label{4trrgiusta}
\alpha_n a_{n+1}+\beta_n a_n+\gamma_n a_{n-1}+\delta_n a_{n-2}=0,
\qquad n\geq 2,
\ee
where:
\beq\label{primostep1}
&&\alpha_n=n(n+1)c_0\\
\nn
&&\beta_n=(n-1)n c_1+n d_0\\
\nn
&&\gamma_n=(n-2)(n-1)c_2+(n-1)d_1+e_0\\
\nn
&&\delta_n=(n-3)(n-2)c_3+(n-2)d_2+e_1
\eeq
The coefficients $a_0$ and $a_1$ can simply be determined by imposing the 
continuity of \op Z_\ell \cl and \op {Z_\ell}_{,r} \cl
in $r=a$, since 
from eq.(\ref{subst}) 
it follows that:
\be\label{a0fc}
a_0=\left.\phi_\ell\right|_{v=0}=\f{Z_\ell(a)}{\chi(a)}
\ee
\be\label{a1fc}
a_1=\left.{\phi_\ell}_{,v}\right|_{v=0}=\f{a}{\chi(a)}
\left[{Z_\ell}_{,r}(a)+\f{i\omega a}{a-1}Z_\ell(a)\right]
\ee
The values  of \op Z_\ell(a) \cl and \op {Z_\ell}_{,r}(a) \cl
can be obtained by numerically integrating 
eq. ( \ref{zz1}) in the interior of the star and continuing the
solution outside, up to \op r=a,\cl by integrating the Regge-Wheeler
equation.
The remaining coefficients can then be determined by recursion from 
eq.(\ref{4trrgiusta}).

To apply the continued fraction technique one has to deal with 
three--term recurrence relations.
Leaver  has shown that the four--term recurrence relation 
(\ref{4trrgiusta})
can be reduced to a three--term relation by a gaussian elimination step
(Leaver, 1990).
In other words, with the positions ($n=0,1$):

\be\label{gausselim1}
\hat{\alpha}_n=\alpha_n,\qquad
\hat{\beta}_n=\beta_n,\qquad
\hat{\gamma}_n=\gamma_n,
\ee
and for $n\geq 2$:

\beq\label{gausselim2}
&&\hat{\alpha}_n=\alpha_n,\qquad
\hat{\beta}_n=\beta_n-\f{\hat{\alpha}_{n-1}\delta_n}{\hat{\gamma}_{n-1}}\\
\nn
&&\hat{\gamma}_n=\gamma_n-\f{\hat{\beta}_{n-1}\delta_n}{\hat{\gamma}_{n-1}},
\qquad \hat{\delta}_n=0,
\eeq
eq.(\ref{4trrgiusta}) reduces to:

\be
\hat{\alpha}_n a_{n+1}+\hat{\beta}_n a_n+\hat{\gamma}_n a_{n-1}=0
\ee
The elimination step is not as trivial as it may seem, 
because in the process one of the {\it three} independent solutions to 
eq.(\ref{4trrgiusta}) is lost.
It can be shown (Leins {\it et al.} 1993) that this solution is not 
relevant for our purposes.

We now turn to investigating the asymptotic behaviour of the coefficients \op a_n \cl in the expansion (\ref{expans}).
Let us make the ansatz:
\be\label{alphan}
\lim_{n\to \infty}\frac{a_{n+1}}{a_n}=1+\frac{h}{n^{1/2}}+\frac{k}{n}+...
\ee
Dividing eq. (\ref{4trrgiusta}) by \op n^2 a_n$, 
keeping terms up to \op\sim n^{-3/2}$ and equating to zero the 
various terms in the expansion in powers of \op n^{-1/2}\cl we find 
the relations:
\beq
&&c_0+c_1+c_2+c_3=0, \\
\nn
&&2c_0+c_1-c_3=0, \\
\nn
&&h^2=2i\omega a, \\
\nn
&&k=-\frac{3}{4}+i\omega (a+1).
\eeq
The first two of these equations are identities. 
Substituting the second pair of equations in \ref{alphan} we get:
\be\label{angrande}
\lim_{n\to \infty} a_n=n^{-3/4+i\omega (a+1)}e^{\pm 2\sqrt{2i\omega a n}}.
\ee
According to a definition given by Gautschi, the solution of eq. 
(\ref{alphan}) corresponding to the plus sign in eq. (\ref{angrande})
is said {\it dominant}, whereas that corresponding to the minus sign  is
said {\it minimal} (Gautschi, 1967). If we select the  minimal solution
we see that the expansion (\ref{expans}) is absolutely 
and uniformly convergent outside the star, provided that we choose 
\op a\cl such that
\be
a/2<R<a \qquad \hbox{and}\qquad a>2,
\ee
and that, according to equation (\ref{subst}),  the solution to equation 
(\ref{mastereq})  behaves as a pure outgoing wave at infinity, 
i.e. it is the wavefunction of a  quasi-normal mode.
Thus, the key point is to identify the  minimal solutions of
eq. (\ref{alphan}).
According to a theorem due to Pincherle (Gautschi 1967), 
if equation (\ref{alphan}) has a minimal solution then the 
following continued fraction relation holds:
\be\label{LNSmatch}
\f{a_1}{a_0}=\f{-\hat{\gamma}_1}{\hat{\beta}_1-}\f{\hat{\alpha}_1\hat{\gamma}_2}{\hat{\beta}_2-}\f{\hat{\alpha}_2\hat{\gamma}_3}{\hat{\beta}_3-}...
\ee
where the continued fraction on the RHS is convergent
and completely determined since the coefficients
\op\hat{\alpha}_n$, \op\hat{\beta}_n$
and \op\hat{\gamma}_n,\cl
defined in eqs. (\ref{gausselim1}),(\ref{gausselim2}) 
are known functions of \op\omega\cl.
Moreover, from eqs. (\ref{a0fc}) and (\ref{a1fc})
it  is apparent that the dependence on the stellar model is all 
contained in the ratio $a_1/a_0$.

Defining now:
\be\label{posizioni}
\hat{\beta}_0=\f{a_1}{a_0}, \qquad
\hat{\alpha}_0=-1,
\ee
eq.(\ref{LNSmatch}) can be recast in the form:
\be\label{cuore}
0=f_0(\omega)=\hat{\beta}_0-\f{\hat{\alpha}_0\hat{\gamma}_1}{\hat{\beta}_1-}\f{\hat{\alpha}_1\hat{\gamma}_2}{\hat{\beta}_2-}\f{\hat{\alpha}_2\hat{\gamma}_3}{\hat{\beta}_3-}...\\
\ee
Using the inversion properties of continued fractions (Wall 1948), 
the latter equation can be inverted \op n \cl times to yield:
\beq\label{cuore2}
0=f_n(\omega)=\hat{\beta}_n-\f{\hat{\alpha}_{n-1}
\hat{\gamma}_n}{\hat{\beta}_{n-1}-}\f{\hat{\alpha}_{n-2}
\hat{\gamma}_{n-1}}{\hat{\beta}_{n-2}-}...\f{\hat{\alpha}_0
\hat{\gamma}_1}{\hat{\beta}_0}- \\
\nn
-\f{\hat{\alpha}_n\hat{\gamma}_{n+1}}{\hat{\beta}_{n+1}-}\f{
\hat{\alpha}_{n+1}\hat{\gamma}_{n+2}}{\hat{\beta}_{n+2}-}
\f{\hat{\alpha}_{n+2}\hat{\gamma}_{n+3}}{\hat{\beta}_{n+3}-}... 
\qquad\hbox{for}\qquad n=1,2,...
\eeq
These \op n \cl conditions are analytically equivalent to eq.(\ref{cuore}). 
Anyway, since the functions \op f_n(\omega) \cl have different convergence 
properties, each of them is best suited to find the quasi--normal 
frequencies in a given region of the complex \op \omega \cl plane.
This is the main reason behind the power and flexibility of the 
continued fraction technique.

To find the mode frequencies we have adopted the following numerical 
procedure. We compute the real and imaginary parts of 
\op f_n(\omega)\cl for a given inversion index \op n \cl (say \op n=0$) 
on a suitably chosen grid of \op (\omega_0,\omega_i) \cl values. 
We plot the curves along which the two functions are zero and look
for the intersections of those curves: 
the quasi--normal frequencies as the points where the 
lines cross.
For each assigned value of the inversion index \op n, \cl besides the 
physical frequencies this method singles out some spurious roots, 
which, however,  can be ruled out easily, since for different values of  
\op n\cl they disappear.   
\begin{figure}
\begin{center}
\leavevmode
\centerline{\epsfig{figure=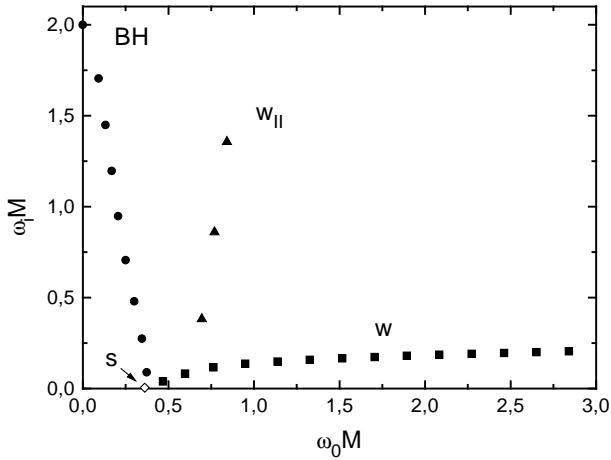,angle=000,width=10cm}}
\caption{The real and the imaginary part of the frequency of the
{w}-modes (square) and of the {w$_{II}$}-modes
 for a homogeneous star with \op R/M=2.4,\cl are compared with the
quasi-normal mode frequencies of a black hole.
The diamond indicates the {\bf s}-mode of the star. }
\label{MODIHOMSTAR}
\end{center}
\end{figure}

The method has been tested by verifying and extending results on 
constant density models of stars previously obtained by Kokkotas (1995), 
who found the QNM frequencies through a direct integration of 
eq.(\ref{zz1}). 
Our results are in excellent agreement (up to the fifth significant digit)
with those of Kokkotas. 
Furthermore, exploiting the inversion properties of continued fraction, 
we were able to find an additional set of modes, 
the so called w$_{II}$-modes (Leins {\it et al.} 1993), 
with still larger imaginary part. The real and imaginary part 
of the axial modes of a homogeneous star with \op R/M=2.4,\cl 
found with the continued fraction method, are 
plotted in fig. \ref{MODIHOMSTAR}  and compared with the
quasi-normal mode frequencies of a Schwarzschild  black hole.

\section{Numerical results}

The continued fraction method has been applied 
to compute the frequency and the damping time of the first 
axial {w}-mode, $\nu_{w_0}$  and
$\tau_{w_0}$, respectively, for models of stars with the EOS's 
discussed in section 3.
In table 1 we summarize the characteristics of the considered
stellar models.
For each EOS, we choose the upper value of the central density 
(column 2) as that of the last radially stable configuration.
The lower value is chosen as in (Andersson \&
Kokkotas 1998) for the EOS's A, B, WFF and L, 
whereas for APR1 and APR2 - that have been
used to compute the oscillation modes of a star in this paper for the
first time  - it is  chosen to give a
value of \op M/R\cl comparable with that of the other models.
In column 3 we give the ratio $M/R$  corresponding to the selected
endpoints of \op\rho_c,\cl 
the radius  and mass  of the
star are given in columns 4 and 5, respectively, and
the values of $\nu_{w_0}$  and
$\tau_{w_0}$ are in columns 6 and 7.

\begin{figure}
\begin{center}
\leavevmode
\centerline{\epsfig{figure=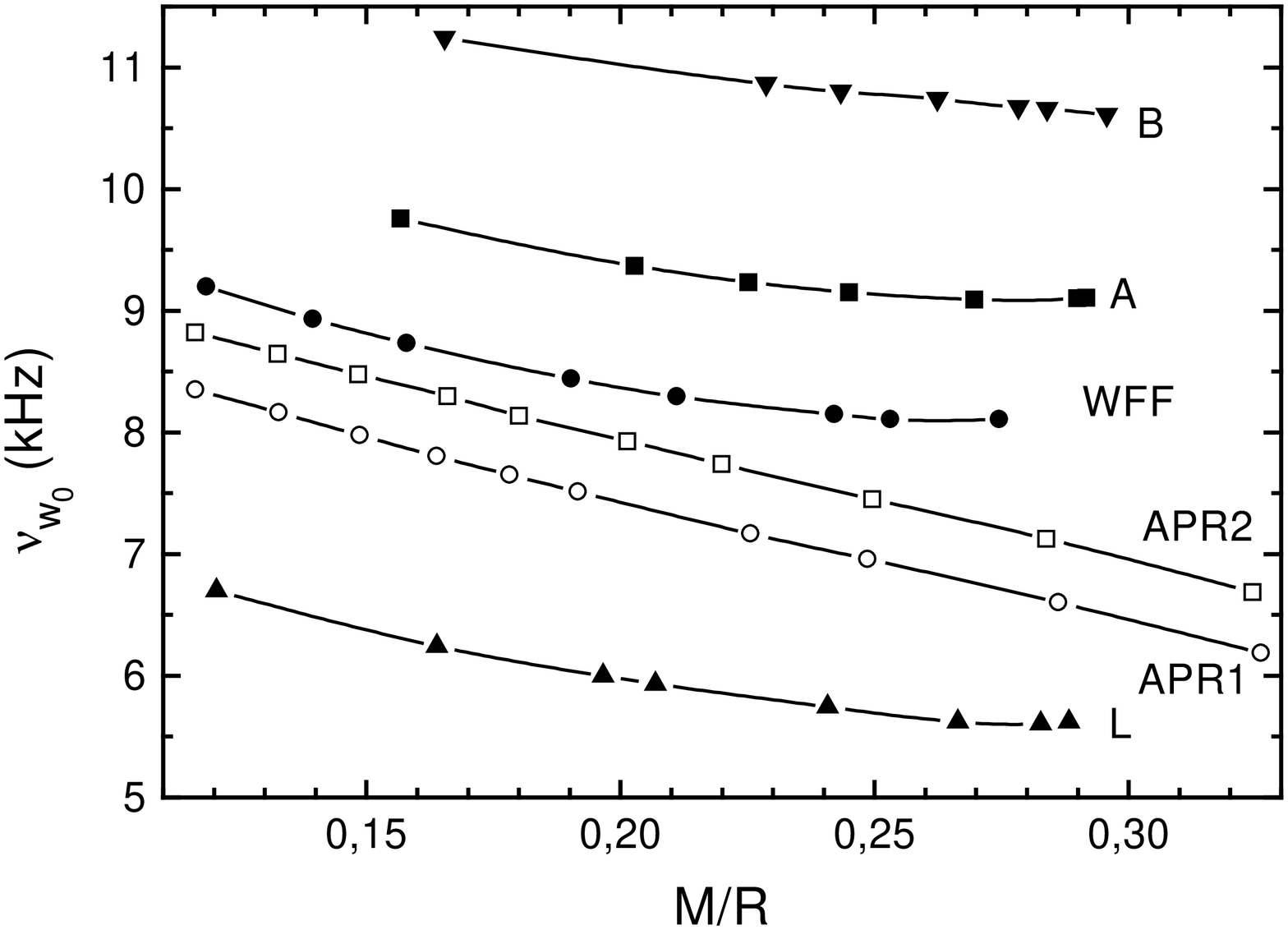,angle=000,width=10cm}}
\caption{The frequency of the first axial {w}-mode is plotted 
as a function of the compactness of the star.}
\label{W0NU}
\end{center}
\end{figure}

In fig.  \ref{W0NU} we plot 
\op \nu_{w_0},\cl as a function of the star compactness,  $ M/R,\cl
for the EOS's given in table 1
and within the corresponding central density range.
The damping time, $ \tau_{w_0}, \cl is similarly  plotted in 
fig. \ref{W0TAU}.
Since the  pressure and density inside the star
are obtained by logarithmically interpolating the pressure-density tables,  
for values of \op M/R\cl close to the stability
limit their reconstructed radial behaviour is  less accurate. 
Hence the corresponding values of 
$\nu_{w_0}$  and $\tau_{w_0}$ are less reliable.
For this reason, we have omitted from 
table 1 and fig. \ref{W0NU} and \ref{W0TAU}
the  marginally stable configurations 
for the WFF equation of state.

From table 1 and fig. \ref{W0NU} we see that, apart from the equations of state 
WFF, APR2 and APR1, for which the ranges within which  \op
\nu_{w_0}\cl  varies partially overlaps, the frequencies of the lower 
{w}-mode for the remaining EOS's, i.e. B, A, WFF and L
range within intervals that are separated. This means that 
if an axial gravitational wave emitted 
by a star  at a frequency of, say, \op 10~kHz\cl could be detected,
then we would be able to identify the equation of state prevailing in 
the star interior as the EOS B (among those we have considered), 
even without knowing the mass and the radius!

\begin{table}
\begin{center}
\begin{tabular}{||l|l|l|l|l|l|l||}
\multicolumn{7}{||c||}{} \\ 
\hline \hline
EOS
&$\rho_c\times 10^{15}$
&$M/R$
&$R$
&$M$
&$\nu$
&$\tau$\\ \hline \hline
B
&$1.995$
&$0.165$
&$8.7$
&$0.97$
&$11.2$
&$20.2$\\
&$5.91$
&$0.296$
&$7.1$
&$1.42$
&$10.6$
&$71.7$\\
\hline
A
&$1.259$
&$0.157$
&$9.9$
&$1.05$
&$9.76$
&$21.6$\\
&$4.11$
&$0.291$
&$8.4$
&$1.65$
&$9.11$
&$72.4$\\
\hline
WFF
&$0.8$
&$0.118$
&$11.1$
&$0.89$
&$9.20$
&$18.6$\\
&$2.6$
&$0.274$
&$9.8$
&$1.83$
&$8.11$
&$62.3$\\
\hline
APR2
&$0.75$
&$0.116$
&$11.8$
&$0.93$
&$8.82$
&$19.4$\\
&$2.75$
&$0.325$
&$10.0$
&$2.20$
&$6.69$
&$165.3$\\
\hline
APR1
&$0.65$
&$0.116$
&$12.4$
&$0.97$
&$8.35$
&$20.4$\\
&$2.36$
&$0.326$
&$10.8$
&$2.38$
&$6.19$
&$177.0$\\
\hline
L
&$0.398$
&$0.120$
&$14.9$
&$1.21$
&$6.70$
&$25.5$\\
&$1.42$
&$0.288$
&$13.6$
&$2.66$
&$5.62$
&$76.7$\\
\hline
\end{tabular}
\end{center}
\caption{
For each EOS, we give the minimum and maximum values 
of the  central density (g$\cdot$cm$^{-3}$) 
chosen as explained in section 5 (column 2).
In correspondence of these endpoints we tabulate: the ratio $M/R$ 
in geometrical units (column 3), the radius (km) and mass ($M_\odot$) of the
star (columns 4 and 5, respectively), the values of $\nu_{w_0}$ (kHz) and
$\tau_{w_0}$ ($\mu$s) of the lowest axial {w}-mode (columns 6 and 7).
}
\end{table}

\begin{figure}
\begin{center}
\leavevmode
\centerline{\epsfig{figure=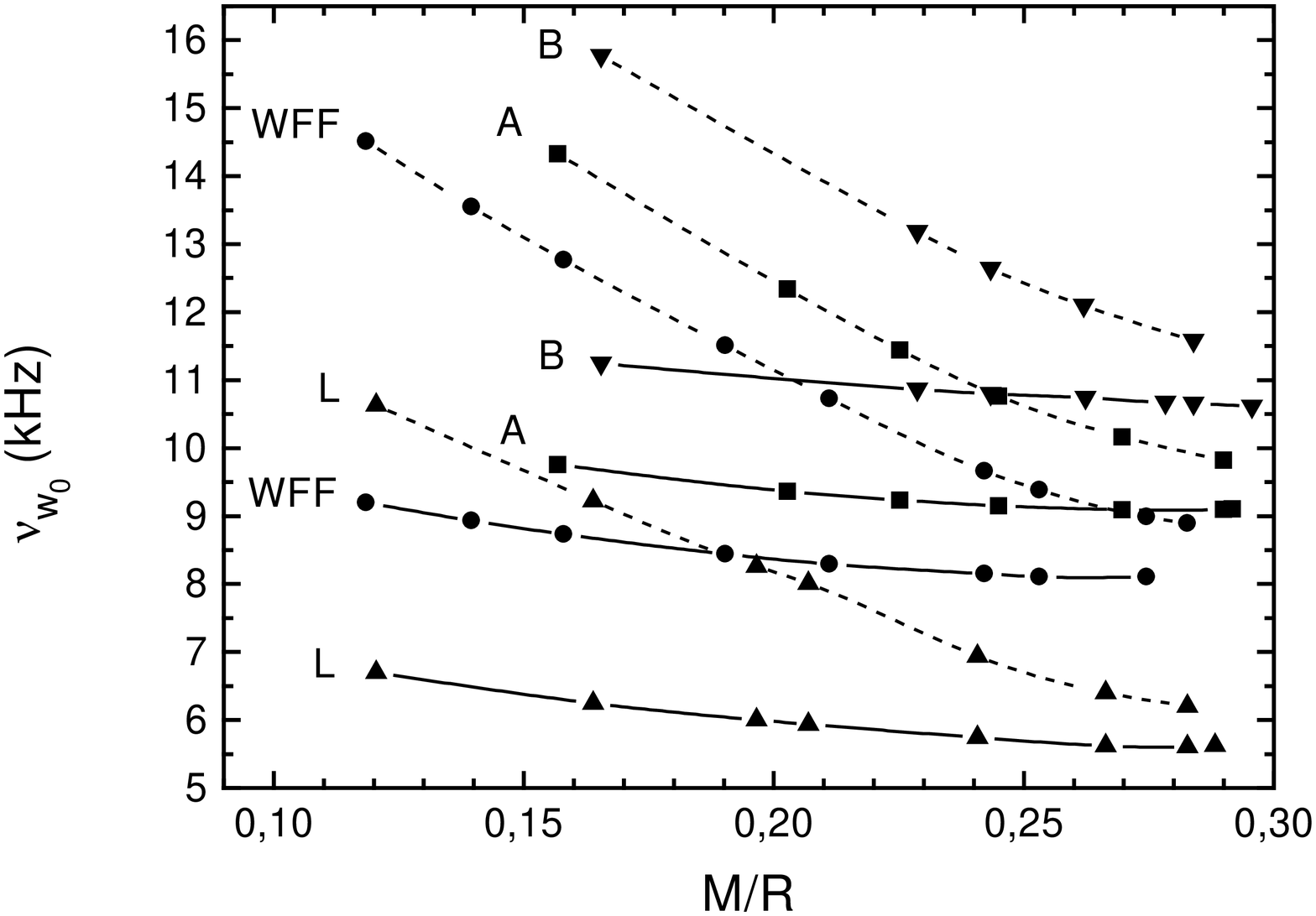,angle=000,width=10cm}}
\caption{The frequency of the first polar (dashed line) and axial
(continuous line) {w}-modes are plotted 
as a function of the compactness of the star for the EOS's A, B, WFF, L.}
\label{W0POLARAXIAL}
\end{center}
\end{figure}

\begin{figure}
\begin{center}
\leavevmode
\centerline{\epsfig{figure=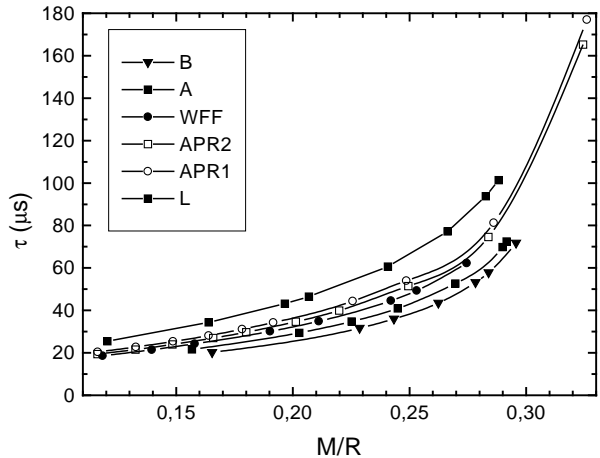,angle=000,width=10cm}}
\caption{The damping time of the first {w}-mode is plotted 
as a function of the compactness of the star.}
\label{W0TAU}
\end{center}
\end{figure}

\begin{table}
\begin{center}
\begin{tabular}{||l|l|l|l|l|l|l||}
\multicolumn{7}{||c||}{} \\ 
\hline \hline
EOS
&$\rho_c\times 10^{15}$
&$M/R$
&$R$
&$M$
&$\nu$
&$\tau$\\ \hline \hline
B
&$1.995$
&$0.165$
&$8.8$
&$0.97$
&$15.8$
&$15$\\
&$5.012$
&$0.283$
&$7.3$
&$1.41$
&$11.6$
&$53$\\
\hline
A
&$1.259$
&$0.157$
&$9.9$
&$1.05$
&$14.3$
&$17$\\
&$3.980$
&$0.290$
&$8.4$
&$1.65$
&$9.8$
&$64$\\
\hline
WFF
&$0.8$
&$0.118$
&$11.1$
&$0.89$
&$14.5$
&$15$\\
&$4.0$
&$0.294$
&$9.2$
&$1.83$
&$8.8$
&$81$\\
\hline
L
&$0.398$
&$0.120$
&$14.9$
&$1.21$
&$10.6$
&$20$\\
&$1.500$
&$0.289$
&$13.6$
&$2.66$
&$6.1$
&$92$\\
\hline
\end{tabular}
\end{center}
\caption{
The central density, radius, mass, and the frequency and the damping time 
of the polar {w}-mode are tabulated, as in table 1 (same units), for four 
equations of state considered both by Andersson and Kokkotas 
and in the present paper.
The data are extracted from the tables A1, A2, A9 and A10
given in  (Andersson \& Kokkotas 1998).
}
\end{table}

In order to check whether a similar identification can be done
by detecting a polar gravitational wave,
it is interesting to compare the frequencies of the fundamental {\it axial} 
{w}-mode with those of the fundamental {\it polar} {w}-mode. 
The latter have been computed by 
Andersson and Kokkotas (Andersson \& Kokkotas 1998) for several equations
of state, including  EOS A, B, WFF and L, which we have also considered.
Using their results,  in fig. \ref{W0POLARAXIAL} we plot the
frequencies of the first {\it axial} and {\it polar} {w}-mode
as a function of the star compactness for these EOS's, and 
in table 2 we tabulate the same quantities given in table 1, and
the  values of the {\it polar} \op \nu_{w_0}\cl and \op \tau_{w_0} \cl 
corresponding to the central density range they consider.
From table 2 and fig. \ref{W0POLARAXIAL} we see that the detection of a 
polar gravitational
wave at a frequency of, say, \op 10~kHz,\cl would not allow to discriminate
between the equations of state A, WFF and L, and similar considerations hold
for  different values of the wave frequency.

Our results show that the frequencies of the axial gravitational waves
emitted by neutron stars oscillating in their {w}-modes
are mainly driven by the stiffness of the EOS, 
which is in turn dictated by the physical assumptions underlying the
model describing the star interior. As a consequence, models WFF, APR1 and
APR2, based on 
very similar assumptions ($\beta$-stable matter with hadronic interactions
constrained by nucleon-nucleon scattering data) yield frequency ranges
that significantly overlap with each other. 

Although the data available on  radiopulsars allow to estimate the mass
of the observed neutron stars, little is known on their radius. It is
therefore useful to provide  methods that give information on this
stellar parameter.
To this purpose, we have fitted the values of 
\op \nu_{w_0}\cl and \op \tau_{w_0} \cl for the six EOS considered in 
this paper, respectively with a linear and a quadratic function of the
compactness \op M/R.\cl
We find the following robust empirical relations
\be
\nu_{w_0} \approx \frac{1}{R}\left[a\f{M}{R}+b\right],
\label{empir1}
\ee
where \op\nu_{w_0}\cl is expressed in kHz, and $R$ in km, with
\beq
&&a=-155.45\pm 3.23,\\\nn
&&b=121.69\pm 0.73,
\eeq
and  $\chi^2=1.84.$ 
The damping time is well approximated by the relation
\be
\f{1}{\tau_{w_0} }\approx 
\frac{1}{\bar M}\left[a\left(\f{M}{R}\right)^2+b\f{M}{R}+c\right],
\label{empir2}
\ee
where \op \tau_{w_0}\cl is in ms, and \op\bar M\cl is the mass expressed in
solar mass units, with 
\beq
&&a=-1672.74\pm 63.39\\\nn
&&b=482.57\pm 27.40\\\nn
&&c=37.37\pm 2.80.
\eeq
and  $\chi^2 = 2.12$.

\section{Concluding Remarks} 

Since the axial perturbations of non rotating stars are described by a
Schr\"odinger equation with a potential barrier shaped by the manner in
which pressure and density are distributed inside the star, the  
frequencies of the axial quasi-normal modes are expected to 
depend on the equation of state. This was shown to happen
also for the polar w-modes (Andersson \& Kokkotas 1998) that are coupled
to negligible fluid motion. In particular, both the  axial and the 
polar w-modes appear to depend essentially on the stiffness of the
equation of state, and this is further
manifested by the possibility of finding a robust fit of both the
frequency and the damping time of these modes as a function of the
compactness of the star. However, for each selected EOS the frequency of
the {\it polar} w-modes is a rather steeply decreasing functions of 
\op M/R,\cl  whereas in the case of 
the {\it axial} ones the dependence of $\nu_{w_0}$ on the star compactness
turns out to be weak.  
This  different behaviour is clearly displayed in fig. \ref{W0POLARAXIAL}, 
where the frequency of  polar and the  axial  w$_0$-mode are 
compared.
As a consequence of this behaviour,  the  axial w-modes give  a
more direct and explicit information, compared to the  polar ones, 
on which equation of state prevails inside the star,
regardless of its mass and  radius. 

We emphasize that 
our results show that EOS's based on different assumptions
correspond to non overlapping frequency ranges. Hence, 
the detection of axial gravitational 
waves would allow to further constrain the existing models, with 
regard to both the composition of neutron star matter and the description 
of the hadronic interactions. 
The real problem is whether we will ever be able to detect an axial
gravitational wave impinging on earth from a ringing star.
Only detailed simulations of astrophysical processes, like the
gravitational collapse or the coalescence of compact bodies, will tell us
to what extent can the w-modes be excited and  be
significant from the point of view of detection. If this will happen to be
the case, since these modes have frequencies higher than those detectable 
by current experiments, new high frequency detectors will need to be planned.

\section*{Acknowledgments}
We would like to thank K.D. Kokkotas for providing tables 
of the pressure-density relations for the EOS's B, A and L, and 
D.J. Ravenhall, who made the results of his calculations available
to us. We also thank P. Nollert and J. Ruoff for their useful suggestions
about the application of the continued fraction method.

\label{lastpage}

\end{document}